\documentclass[aps,pra,reprint,groupedaddress,showpacs]{revtex4-1}
\usepackage{amssymb}
\usepackage{amsmath}
\usepackage{dcolumn}
\usepackage{bm}
\usepackage{graphicx}
\usepackage{mathrsfs}

\begin{document}

\title{Engineering two-mode entangled states between two superconducting resonators by dissipation}
\author{Peng-Bo Li}
\email{lipengbo@mail.xjtu.edu.cn}
\author{Shao-Yan Gao}
\email{gaosy@mail.xjtu.edu.cn}
\author{Fu-Li Li}
\affiliation {MOE Key Laboratory for Nonequilibrium Synthesis and Modulation of Condensed Matter,\\
Department of Applied Physics, Xi'an Jiaotong University, Xi'an
710049, China}

\begin{abstract}
We present an experimental feasible scheme to synthesize two-mode continuous-variable entangled states of two superconducting resonators that are interconnected by two gap-tunable superconducting qubits. We show that, with each artificial atom suitably driven by a bichromatic microwave field to induce sidebands in the qubit-resonator coupling, the stationary state of the photon fields in the two resonators can be cooled and steered into a two-mode squeezed vacuum state via a dissipative quantum dynamical process, while the superconducting qubits remain in their ground states.
In this scheme the qubit decay plays a positive role and can help drive the system to the target state, which thus converts a detrimental source of noise into a
resource.

\end{abstract}
\maketitle

\section{introduction}

Solid state devices based on superconducting circuits
are promising candidates for studying fundamental physics and implementing quantum information protocols \cite{pic,nature-453-1031,RMP-73-357,low-Tem-Phys,pt-1}.
These devices, which can be designed and
fabricated on demand, provide an unprecedented
level of tunability and flexibility in the implementation of scalable quantum information processing.
Three basic types of superconducting qubits, including charge\cite{nature-398-786}, flux\cite{sci-285-1036} and phase qubits\cite{sci-296-889}, have been theoretically and experimentally
investigated to realize quantum computation and communication. In general, these qubits are coupled either through circuit elements such as capacitances  and inductances,  or via a quantum bus \cite{nature-451-664,nature-449-443}, a superconducting resonator. However, recent researches have shown that superconducting resonators can also be addressed and manipulated for quantum state generation \cite{nature-454-310,prl-101-240401,prl-101-253602,prl-95-140504,prb-76-205416,prb-70-205304,prb-75-184508,pra-83-035807}. Particularly, plenty of works have focused on the extension
from single to more versatile multi-resonator architectures \cite{prb-78-104508,prb-81-085328,pra-79-042339,epl-85-50007}, where the resonators are interconnected by superconducting qubits.
These multi-resonator systems allow manipulation of spatially separated photon modes and engineering quantum states between physically distant
cavities \cite{Nature-physics-7-287,prl-106-060401,prl-105-050501,prl-106-257002,prb-80-214538,prb-76-064305,prb-85-024537,njp-12-093036,pra-79-063811,pra-73-022318,pra-85-014303,pra-85-022324,pra-83-022302}.

In this work, we introduce an efficient scheme for the preparation of two-mode entangled states between two spatially separated superconducting
resonators that are interconnected by two gap-tunable superconducting qubits. This proposal exploits a bichromatic microwave driving of each qubit
to modify its coupling to the resonator modes, inducing sidebands in the qubit-resonator coupling \cite{prb-76-144518,prl-108-043602},  and actively utilizes the qubit decay to drive the system to the desired state. We show that, combining the sidebands with the qubit decay, the stationary state of the photon fields in the two resonators can be steered into a two-mode squeezed vacuum state via a dissipative quantum dynamical process, while the superconducting qubits remain in their ground states. We detail the actual physical implementation of this scheme with the superconducting LC resonators inductively coupled to
two gap-tunable $\alpha$-loop flux qubits. This work may open up promising perspectives for quantum information processing with coupled networks of
superconducting resonators.

\section{The model}
\begin{figure}[h]
\centerline{\includegraphics[bb=93 583 522 774,totalheight=1.1in,clip]{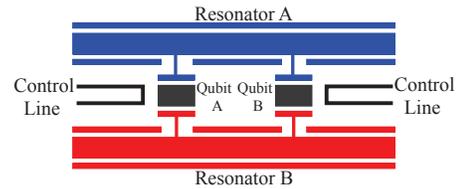}}
\caption{(Color online) The schematic for generating entanglement between two superconducting resonators.
These resonators are interconnected by two gap-tunable superconducting qubits, which are independently controlled by external circuits.}
\end{figure}
We consider two gap-tunable superconducting qubits with ground state $\vert g\rangle$ and excited state $\vert e\rangle$, coupling to two superconducting resonators, as shown in Fig. 1. The energy levels of each qubit is individually driving by a bichromatic microwave field to induce sidebands in the qubit-cavity coupling. A typical gap-tunable superconducting qubit is the modified flux qubit where two junctions forms a superconducting quantum interference device (SQUID) to allow the modulation of the qubit gap \cite{prl-102-090501}. This modified flux qubit allows a fast change of the qubit resonance
frequency while remaining at the degeneracy point.
The free Hamiltonian describing the two qubits and the resonator modes is (let $\hbar=1$)
\begin{equation}\label{H1}
   \hat{ H}_0=\sum_{\lambda=1}^2(\frac{\delta}{2}\sigma_z^\lambda+\nu_\lambda\hat{a}_\lambda^\dag\hat{a}_\lambda),
\end{equation}
where $\delta$ is the qubit gap without external driving, $\sigma_z^\lambda$ is the Pauli operator, $\nu_\lambda$ is the resonator frequency, and $\hat{a}_\lambda^\dag$ is the creation operator for the $\lambda$\emph{th} resonator. The bare qubit-resonator interaction is given by
\begin{equation}\label{H2}
    \hat{H}_I=\sum_{\lambda=1}^2\sum_{l=1}^2g_l^{(\lambda)}(\sigma_+^\lambda+\sigma_-^\lambda)(\hat{a}_l^\dag+\hat{a}_l),
\end{equation}
where $g_l^{(\lambda)}$ is the coupling strength between the qubits and the resonators, and $\sigma_{\pm}^j$ the qubit ladder operators.
We now consider a $\sigma_z$ driving with frequencies $\omega^{(\lambda)}_{d,l}$ for the $\lambda$\emph{th} qubit \cite{prl-108-043602}. For flux qubits, this driving can be implemented by
an external magnetic field that induces a flux driving \cite{prl-102-090501}. Expressing the driving amplitude by the ratios $\xi_l^{(\lambda)}$ to the driving frequencies, we have
\begin{equation}\label{H3}
    \hat{H}_d(t)=-\sum_{\lambda=1}^2\sum_{l=1}^2\xi_l^{(\lambda)}\omega^{(\lambda)}_{d,l}\cos(\omega^{(\lambda)}_{d,l}t)\sigma_z^\lambda.
\end{equation}

In the following, the system-environment interaction is
assumed Markovian, and then is described by a master
equation in Lindblad form. Here we consider only relaxation and dephasing of the superconducting
qubits, with the relaxation rate
$\Gamma_r$ and dephasing rate $\Gamma_\varphi$. The
time evolution of the density operator $\hat{\rho}$ for the whole system is then described by the
master equation
\begin{eqnarray}
\label{M1}
\frac{d\hat{\rho}}{dt}&=&-i[\hat{H},\hat{\rho}]+\Gamma_r D[\sigma_-^1]\hat{\rho}+\Gamma_r D[\sigma_-^2]\hat{\rho}\nonumber\\
&&+\Gamma_{\varphi}D[\sigma_z^1]\hat{\rho}+\Gamma_{\varphi}D[\sigma_z^2]\hat{\rho},
\end{eqnarray}
where $\hat{H}=\hat{H}_0+\hat{H}_d+\hat{H}_I$, and $D[A]\hat{\rho}=2A\hat{\rho}A^\dag-A^\dag A\hat{\rho}-\hat{\rho}A^\dag A$.

We proceed to perform the unitary transformation $U_1(t)=e^{-i\hat{H}_0t}$,
which leads to
\begin{eqnarray}
\label{H4}
\hat{H}&=&\sum_{\lambda=1}^2\sum_{l=1}^2g_l^{(\lambda)}(e^{i\delta t}\sigma_+^\lambda+e^{-i\delta t}\sigma_-^\lambda)(e^{i\nu_l t}\hat{a}_l^\dag+e^{-i\nu_l t}\hat{a}_l)\nonumber\\
&&-\sum_{\lambda=1}^2\sum_{l=1}^2\xi_l^{(\lambda)}\omega^{(\lambda)}_{d,l}\cos(\omega^{(\lambda)}_{d,l}t)\sigma_z^\lambda.
\end{eqnarray}
To engineer the desired qubit-photon coupling, we perform another unitary transformation
\begin{eqnarray}
\label{U}
U_2(t)&=&Te^{-i\int_0^t\hat{H}_d(\tau)d\tau}\nonumber\\
&=&1-i\int_0^t\hat{H}_d(t_1)dt_1\nonumber\\
&&+(-i)^2\int_0^tdt_1\int_0^{t_1}dt_2\hat{H}_d(t_1)\hat{H}_d(t_2)+...
\end{eqnarray}
Keeping only the leading order in the rations $\xi_l^{(\lambda)}$, we have
\begin{eqnarray}
\label{H5}
\hat{H}&=&\sum_{\lambda=1}^2\sum_{j=1}^2\sigma_+^\lambda\{g_j^{(\lambda)} e^{i\delta t}(e^{i\nu_j t}\hat{a}_j^\dag+e^{-i\nu_j t}\hat{a}_j)\nonumber\\
&&[1-\sum_{l=1}^2\xi_l^{(\lambda)}(e^{i\omega^{(\lambda)}_{d,l}t}-\text{h.c})]
\}+\text{H.c.}
\end{eqnarray}
By setting $\omega^{(1)}_{d,1}=\delta-\nu_1,\omega^{(1)}_{d,2}=\delta+\nu_2$, $\omega^{(2)}_{d,1}=\delta+\nu_1,\omega^{(2)}_{d,2}=\delta-\nu_2$, and
neglecting the fast oscillating terms under the rotating-wave approximation, we can get
\begin{eqnarray}
\label{H6}
\hat{H}&=&\sigma_+^1(g_1^{(1)}\xi_1^{(1)}\hat{a}_1+g_2^{(1)}\xi_2^{(1)}\hat{a}_2^\dag)+\sigma_+^2(g_1^{(2)}\xi_1^{(2)}\hat{a}_1^\dag+g_2^{(2)}\xi_2^{(2)}\hat{a}_2)\nonumber\\
&&+\text{H.c.}
\end{eqnarray}
This Hamiltonian describes the red sideband ($\omega^{(1)}_{d,1}=\delta-\nu_1$)
and blue sideband ($\omega^{(1)}_{d,2}=\delta+\nu_2$) excitations of the
first qubit induced by the bichromatic microwave field with frequencies $\omega^{(1)}_{d,1}$ and $\omega^{(1)}_{d,2}$, and
the blue sideband ($\omega^{(2)}_{d,1}=\delta+\nu_1$)
and red sideband ($\omega^{(2)}_{d,2}=\delta-\nu_2$) excitations of the
second qubit induced by another bichromatic microwave field with frequencies $\omega^{(2)}_{d,1}$ and $\omega^{(2)}_{d,2}$.
The schematic diagram of these engineered qubit-photon couplings is shown in Fig. 2.
\begin{figure}[h]
\centerline{\includegraphics[bb=161 550 423 730,totalheight=1.5in,clip]{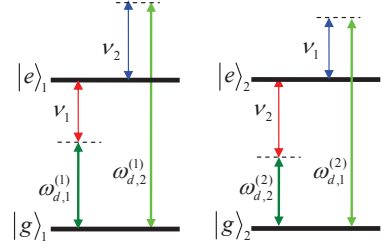}}
\caption{(Color online) The schematic of the engineered sideband couplings for the two superconducting qubits. The energy gap of the $j$th
qubit oscillates at frequencies $\omega^{(j)}_{d,1}$ and $\omega^{(j)}_{d,2}$ induced by the flux driving of $\phi^{(j)}(t) (j=1,2)$.}
\end{figure}

We choose $g_1^{(1)}\xi_1^{(1)}=g_2^{(2)}\xi_2^{(2)}=\Theta_1$, $g_2^{(1)}\xi_2^{(1)}=g_1^{(2)}\xi_1^{(2)}=\Theta_2$, and $\Theta_1>\Theta_2$. In this case the Hamiltonian (\ref{H6}) can be rewritten as
\begin{eqnarray}
\label{H7}
\hat{H}&=&\sigma_+^1(\Theta_1\hat{a}_1+\Theta_2\hat{a}_2^\dag)+\sigma_+^2(\Theta_1\hat{a}_2+\Theta_2\hat{a}_1^\dag)
+\text{H.c.}
\end{eqnarray}
We apply a unitary transformation $\tilde{\rho}=\hat{S}^\dag(\zeta)\hat{\rho}\hat{S}(\zeta)$, where $\hat{S}(\zeta)=e^{\zeta\hat{a}_1\hat{a}_2-\zeta\hat{a}_1^\dag\hat{a}_2^\dag},\zeta=\tanh^{-1}\Theta_2/\Theta_1$. The master equation (\ref{M1}) for the system now becomes
\begin{eqnarray}
\label{M2}
\frac{d\tilde{\rho}}{dt}&=&-i[\tilde{H},\tilde{\rho}]+\Gamma_r D[\sigma_-^1]\tilde{\rho}+\Gamma_r D[\sigma_-^2]\tilde{\rho}\nonumber\\
&&+\Gamma_{\varphi}D[\sigma_z^1]\tilde{\rho}+\Gamma_{\varphi}D[\sigma_z^2]\tilde{\rho},
\end{eqnarray}
where
\begin{eqnarray}
\label{H8}
\tilde{H}&=&\sqrt{\Theta^2_1-\Theta^2_2}(\hat{a}_1\sigma_+^1+\hat{a}_2\sigma_+^2)+\text{H.c}.
\end{eqnarray}

The master equation (\ref{M2}) is equivalent to that of two independent Jaynes-Cummings models of two two-level atoms coupled to two cavity modes respectively, with atomic
spontaneous emission and dephasing included.
In the transformed picture, the $j$th superconducting qubit pulls the $j$th resonator mode into the vacuum state by a process
in which the qubit continuously absorbs photons and decays back to its ground state.
Therefore,  the steady state of the system in the transformed picture is the vacuum $\vert00\rangle_{r}$ for the resonator modes $\hat{a}_1,\hat{a}_2$, and $\vert g\rangle_1\vert g\rangle_2$ for the superconducting qubits.
Reversing the unitary transformation,
it can readily be seen that the state
\begin{eqnarray}
\label{SS}
\vert \psi_S\rangle&=&\hat{S}(\zeta)\vert00\rangle_{r}\vert g\rangle_1\vert g\rangle_2
\end{eqnarray}
is the unique stationary state of the master equation (\ref{M1}) i.e.,
\begin{eqnarray}
\hat{\rho}(t\rightarrow\infty)&=&\vert \psi_S\rangle\langle \psi_S\vert.
\end{eqnarray}

The steady state (\ref{SS}) does
not depend on the initial photon state of the
resonators. Thus the resonators does not have to be initially cooled to the vacuum
state in order to prepare such a state.
Another distinct feature is that, in this scheme the qubit decay plays a positive role and can help drive the system to the target state, which thus converts a detrimental source of noise into a resource. This  feature is  particularly favorable for superconducting quantum circuits,
since decoherence of the superconducting qubits is one of the generic sources of noise and sets the limits of quantum coherence in superconducting quantum circuits. The state $e^{\zeta\hat{a}_1\hat{a}_2-\zeta\hat{a}_1^\dag\hat{a}_2^\dag}\vert00\rangle_r$ is a two-mode squeezed state of the photon fields in the two resonators, which
exhibits Einstein-Podolsky-Rosen (EPR) entanglement.
The degree of squeezing (squeeze parameter $\zeta$) is determined by the ratio of $\Theta_2$ to $\Theta_1$, which can be controlled on demand through tuning the external experimental parameters.

To check the above model, we proceed to numerically
solve the master equation including the decay of the resonator modes, i.e.,
\begin{eqnarray}
\label{M3}
\frac{d\hat{\rho}}{dt}&=&-i[\hat{H},\hat{\rho}]+\Gamma_r D[\sigma_-^1]\hat{\rho}+\Gamma_r D[\sigma_-^2]\hat{\rho}+\Gamma_{\varphi}D[\sigma_z^1]\hat{\rho}\nonumber\\
&&+\Gamma_{\varphi}D[\sigma_z^2]\hat{\rho}+\kappa_1 D[\hat{a}_1]\hat{\rho}+\kappa_2 D[\hat{a}_2]\hat{\rho},
\end{eqnarray}
where $\kappa_\lambda$ is the photon loss
rate for the $\lambda$th resonator.
To quantify the validity of the proposal, we exploit the total variance $V=\langle(\Delta \hat{u})^2+(\Delta \hat{v})^2\rangle$ of a pair of EPR-like operators $\hat{u}=\hat{X}_1+\hat{X}_2$, and  $\hat{v}=\hat{P}_1-\hat{P}_2$ , with $\hat{X}_\lambda=(\hat{a}_\lambda+\hat{a}^\dag_\lambda)/\sqrt{2}$, and $\hat{P}_\lambda=-i(\hat{a}_\lambda-\hat{a}^\dag_\lambda)/\sqrt{2},\lambda=1,2$. A two-mode
Gaussian state is entangled if and only if $V<2$. For an ideal two-mode squeezed vacuum state, the total variance $V=\langle(\Delta \hat{u})^2+(\Delta \hat{v})^2\rangle=2e^{-2\zeta}$, implying this state exhibits EPR entanglement.

In Figure 3 we display the numerical results for time evolution of the total
variance $V$ together with the results for an ideal two-mode  squeezed vacuum state,
under different values for the decay rates $\Gamma$, and $\kappa$, where $\Gamma_r\simeq\Gamma_\varphi\simeq\Gamma$, and $\kappa_1\simeq\kappa_2\simeq\kappa$ are assumed. The relevant parameters
are chosen such that they are within the parameter range for
which this scheme is valid and are accessible with present-day experimental setups. The initial state of the
system is chosen as $\hat{\rho}(0)=\vert00\rangle_r\vert g\rangle_1\vert g\rangle_2$. From the simulation one can readily see that the model
performs very well with the chosen parameters. At steady state ideal EPR entanglement ($\Theta_2/\Theta_1=0.75,V=0.286$) between the photon fields in the two resonators
has been established if the resonator decay rate satisfies $\kappa< \{\Theta_1,\Theta_2,\Gamma\}$. However, when the resonator decay is comparable to
or larger than $\Gamma$, the ideal EPR entanglement is spoiled by resonator decay. Therefore, implementing this
proposal with high fidelity requires that $\kappa<\{\Theta_1,\Theta_2,\Gamma\}.$

\begin{figure}[h]
\centerline{\includegraphics[bb=23 201 596 712,totalheight=2in,clip]{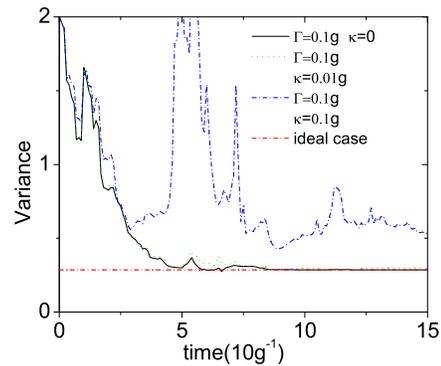}}
\caption{(Color online) Time evolution of the total
variance $V$ through numerically solving the master equation (14), together with
the result for an ideal two-mode squeezed vacuum state. The relevant
parameters are chosen as $\xi_1^{(1)}\simeq\xi_2^{(2)}\simeq\xi_1=0.2,\xi_2^{(1)}\simeq\xi_1^{(2)}\simeq\xi_2=0.15,g_1^{(1)}\simeq g_2^{(1)}\simeq g_1^{(2)} \simeq
g_2^{(2)}=g$.}
\end{figure}

\section{Two-mode entanglement between two superconducting LC resonators }

In principle, the generic model can be implemented with various superconducting resonators coupled by two Josephson-junction-based
superconducting qubits. Here we propose a convenient demonstration with two superconducting LC resonators inductively coupled to two
gap-tunable  flux qubits, as shown in Fig. 4.

The superconducting LC resonators made of a capacitor and an inductor are described
by a simple harmonic oscillator Hamiltonian \cite{prl-96-127006,prl-105-237001} $\hat{H}_r=\sum_{\lambda=1}^2\nu_\lambda(\hat{a}_\lambda^\dag\hat{a}_\lambda+1/2)$, where the
resonance frequency $\nu_\lambda$ is determined by the respective capacitance $C_\lambda$
 and the inductance $L_\lambda$: $\nu_\lambda=1/\sqrt{L_\lambda C_\lambda}$. For a micrometer scale LC resonator, the resonance frequency $\nu_\lambda$ is on the order of several GHz\cite{prl-96-127006,prl-105-237001}.
 The quality factor $Q$ for a LC resonator can reach $Q\geq 10^3$, which leads to a decay rate for the resonator on the order of several MHz.
 These LC resonators are
 interconnected by two gap-tunable flux qubits which are inductively coupled to the resonators.

\begin{figure}[h]
\centerline{\includegraphics[bb=166 473 459 740,totalheight=1.8in,clip]{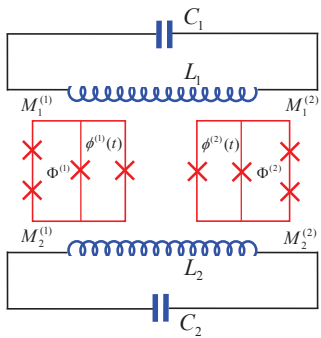}}
\caption{(Color online) The schematic of two superconducting LC resonators inductively coupled to the
$\alpha$-loop flux qubits. The $\lambda$th flux qubit, controlled by the static magnetic flux $\Phi^{(\lambda)}$ and time-dependent flux driving
$\phi^{(\lambda)}(t)$, is coupled to the $l$th resonator via mutual inductance $M^{(\lambda)}_l$. }
\end{figure}
We consider the so-called tunable-$\alpha$ flux qubits, as demonstrated in Ref. \onlinecite{prl-102-090501}.
Different from the flux qubit with three Josephson junctions, one of which has the coupling energy  smaller
than that of the other two junctions by a factor $\alpha$, here the small $\alpha$
junction is replaced by a so-called $\alpha$ loop, formed by a SQUID
with two identical Josephson junctions.
The $\lambda$th flux qubit can be  operated at the degeneracy point with an external applied magnetic flux of $\Phi^{(\lambda)}\simeq \Phi_0/2$ and behaves
effectively as a two-level system, where $\Phi_0=h/2e$ is the flux quantum.  The qubit gap can be controlled via the flux driving $\phi^{(\lambda)}(t)$ through
the SQUID loop, which can be implemented by the external microwave control lines.
In this case, the Hamiltonian for the $\lambda$th flux qubit can be written as
\begin{equation}
    \hat{H}_q^{(\lambda)}=-\frac{1}{2}\{\epsilon \sigma_z^{(\lambda)}+\Delta[\phi^{(\lambda)}(t)]\sigma_x^{(\lambda)}\}
\end{equation}
in the basis of the persistent current states $\{\vert \circlearrowright\rangle,\vert \circlearrowleft\rangle\}$.
Here $\epsilon=2I_p(\Phi^{(\lambda)}-\Phi_0/2)$, with $I_p$ is the persistent current in the qubit loop. The qubit gap $\Delta[\phi^{(\lambda)}(t)]$ depends on the flux driving, and can be separated into a static part $\delta$ and a time-dependent oscillating part $\propto \cos(\omega^{(\lambda)}_{d,l}t) $.

The two superconducting LC resonators couple to the flux qubits via the mutual inductance. The interaction between the
$\lambda$th qubit and the $l$th LC resonator can be described by a coupling of dipolar nature in the basis of the persistent current states
\begin{equation}
 \hat{H}^{(\lambda l)}_I=g^{(\lambda)}_l\sigma^{(\lambda)}_z(\hat{a}_l^\dag+\hat{a}_l),
\end{equation}
where the strength of the coupling is $g^{(\lambda)}_l=M^{(\lambda)}_lI_p\sqrt{\frac{\nu_l}{2L_l}}$, with $M^{(\lambda)}_l$ the mutual inductance between the $\lambda$th qubit and the $l$th LC resonator. In the basis of the eigenstates of the qubits, $\{\vert g\rangle,\vert e\rangle\}$, the
total Hamiltonian reads
\begin{eqnarray}\label{H}
\hat{H}&=&\sum_{\lambda=1}^2[\frac{\omega_q}{2}\sigma_z^\lambda+\nu_\lambda\hat{a}_\lambda^\dag\hat{a}_\lambda\nonumber\\ &&+\sum_{l=1}^2g_l^{(\lambda)}(\cos\theta\sigma_z^\lambda-\sin\theta\sigma_x^\lambda)(\hat{a}_l^\dag+\hat{a}_l)]
\end{eqnarray}
with $\omega_q=\sqrt{\epsilon^2+\Delta^2}$ and $\tan\theta=\Delta/\epsilon$. At the degeneracy point $\epsilon\simeq0$, this leads to
\begin{eqnarray}\label{TH}
\hat{H}&=&\sum_{\lambda=1}^2\{[\frac{\delta}{2}-\sum_{l=1}^2\xi_l^{(\lambda)}\omega^{(\lambda)}_{d,l}\cos(\omega^{(\lambda)}_{d,l}t)]\sigma_z^\lambda+\nu_\lambda\hat{a}_\lambda^\dag\hat{a}_\lambda\nonumber\\ &&+\sum_{l=1}^2g^{(\lambda)}_l(\sigma_+^\lambda+\sigma_-^\lambda)(\hat{a}_l^\dag+\hat{a}_l)\}
\end{eqnarray}
From this Hamiltonian, one can engineer the desired photon-qubit couplings and prepare the photon fields in the
two LC resonators into the two-mode squeezed vacuum state, via a dissipative quantum dynamical process following the
reasoning in the above section.

We now discuss the relevant experimental parameters. Taking the capacitance and inductance of the LC resonators as
12 pF and 250 pH leads to $\nu_1\simeq\nu_2\simeq2.9$ GHz \cite{nature-431-159}.  The mutual inductance $M^{(\lambda)}_l$ between the $\lambda$th flux qubit and the $l$th LC resonator can be about 20 pH \cite{nature-431-159}. Thus we can obtain the coupling strength $g\simeq 200$ MHz. If we take
$\xi_1=0.2,\xi_2=0.15$, then we obtain the effective coupling strength $\Theta_1\simeq40$ MHz, and $\Theta_2\simeq30$ MHz. The time  for preparing the steady state (\ref{SS}) is determined by the decay rate $\Gamma$ of the flux qubit.  Provided that $\sqrt{\Theta_1^2-\Theta_2^2}\geq \Gamma$,
this time will be the order of a few times $1/\Gamma$. Moreover, if $\kappa\leq \Gamma$, the generated state will be nearly an ideal
two-mode squeezed vacuum state.
If we take $\Gamma\simeq 20$ MHz for the flux qubit, the preparing time will be about 250 ns.

\section{conclusion}
To conclude, we have presented an efficient scheme for the preparation of entangled states between two superconducting resonators that are interconnected by two gap-tunable superconducting qubits.
We have shown that, with each qubit individually driven by a bichromatic microwave field to induce sidebands in the qubit-cavity coupling,
the stationary state of the photon fields in the two resonators can be steered into a two-mode squeezed vacuum state via a dissipative quantum dynamical process. This proposal actively exploits the qubit decay to drive the
system to the desired state and does not depend on the initial photon state of the
resonators, which can be implemented with superconducting LC resonators inductively coupled to $\alpha$-loop flux qubits.

\section*{acknowledgement}
This work is supported by
the NNSF of China under Grants No. 11104215, the Special Prophase Project in the National Basic
Research Program of China under Grant No. 2011CB311807,
and the Research Fund for the Doctoral Program of Higher
Education of China under Grant No. 20110201120035. S.-Y.G.
acknowledges financial support from the Natural Science
Basic Research Plan in the Shaanxi Province of China
(No. 2010JQ1004).

%\bibliography{paper17}

\end{document}